# Geological interpretation of Mount Ciremai geothermal system from remote sensing and magneto-teluric analysis

Prihadi Sumintadireja[1], Asep Saepuloh[1], Dasapta E. Irawan[1*], Diky Irawan[2], and A. Fadillah[3]

[1]Applied Geology Research Division, Faculty of Earth Sciences and Technology, InstitutTeknologi

Bandung

[2]Volcanology and Geothermal Sub Laboratory, Faculty of Earth Sciences and Technology, InstitutTeknologi Bandung

[3]Dinas Energi dan Sumber Sumber Daya Mineral, Provinsi Jawa Barat

Corresponding author: Dasapta E. Irawan, erwin@fitb.itb.ac.id

**Abstract.** The exploration of geothermal system at Mount Ciremai has been started since the early 1980's and has just been studied carefully since the early 2000's. Previous studies have detected the potential of geothermal system and also the groundwater mechanism feeding the system. This paper will discuss the geothermal exploration based on regional scale surface temperature analysis with Landsat image to have a more detail interpretation of the geological setting and magneto-telluric (MT) survey at prospect zones (which identified by the previous method), to have a more exact and in depth local scale structural interpretation. Both methods are directed to pin point appropriate locations for geothermal pilot hole drilling and testing.

We used four scenes of Landsat Enhanced Thematic Mapper (ETM+) data to estimate the surface manifestation of a geothermal system. Temporal analysis of Land Surface Temperature (LST) was applied and coupled with field temperature measurement at seven locations. By combining the TTM with NDVI threshold, the authors can identify six zones with surface temperature anomaly. Among these six zones, three zones were interpreted to have a relation with geothermal system and the other three zones were related to human activities. Then, MT survey was performed at the three geothermal prospects identified from previous remote sensing analysis, at the east flank of the volcano, to estimate the detail subsurface structures. The MT survey successfully identified four buried normal faults in the area, which positively are a part of the conduits in the geothermal system east of Mount Ciremai. From MT analysis, the author also found the locations of volcanic zone, bedrock zone, and the prospect zone. The combination of Landsat analysis on regional scale and MT measurement on a more detail scale has proven to be the reliable method to map geothermal prospect area.

# 1 Introduction

Indonesia is known as one of the country, which lies on the volcanic ring of fire, with the sum of 130 active volcanoes. Carranza *et al*. (2008) has proposed a conceptual modeling and predictive mapping of potential geothermal resources at the regional-scale data sets in West Java, based on the analysis of the spatial distribution of geothermal prospects and thermal springs. The author has mentioned that high potential geothermal zones are occupying 25% of West Java, solely. Ironically, such vast potential of geothermal energy is not maximally extracted due to the series of government policy, which still strongly rely on fossil fuels. However, recently, the role of geothermal energy in Indonesia has been rising. A number of new fields have been explored, including the ones that located on the frontier areas. Irawan *et al*. (2009), Herdianita *et al*. (2010), and Isa *et al*. (2011) have proposed that more regional methods must be developed and a more detail subsurface map have to be built to understand geothermal system. Regional methods like satellite imagery analysis and detail MT measurements are strongly needed to accompany hydrochemical analysis in prospect areas. However, Smirnov (2004) and Gamble *et al*. (1979) also mentioned the limitations of Landsat image and MT analysis. According to those papers, Landsat image analysis was strongly confined by cloud intensity. Then MT analysis must be supported by field observation and other geophysical approach to seek for similar structural pattern on the result.

Mount Ciremai is one of the new locations that have been on the government's list of areas to be explored and opened for the geothermal operator market (Figure 1). Currently, the location is classified as green fields with several manifestations that must be carefully studied before the authorities can put it into the potential field-offering list. In this paper the author will discuss about the application of remote sensing and MT survey in geothermal exploration at Mount Ciremai.

Remote sensing and MT were applied in this study complementary to the field geological observation. Both methods are proven very useful in geothermal prospecting campaign (Kratt *et al*., 2006; Sumintadireja *et al*., 2010). Remote sensing is very handy in mapping large area in an efficient manner in term of scale since it covers the hardly explored volcanic terrain (Carranza *et al*., 2008). Kratt *et al*. (2006) have demonstrated the effectiveness of using Advanced Spaceborne Thermal and Emitted Reflectance Radiometer (ASTER) imagery for mapping specific minerals (borate) in a geothermal field. The image can be used to detect minerals using their ability to sense reflectance value in every pixel. Similar technique also use by many researchers, one of the first whom use the technique was Hodder (1970). Hodder used the visible and near-infrared wave to delineate anomalous spectral reflectance associated with hydrothermal alteration minerals. Moreover, the decreasing of satellite image cost and more images freely distributed in public domain agreement are also one of the advantages in using this technique. Then, we will use the potential locations from remote sensing analysis, as a guide to choose the location for the MT measurement.

Magnetotelluric (MT) has been used extensively in geothermal exploration. Several examples of MT application have been published by Uchida and Sasaki (2006), de Lugao *et al*. (2002), and Pellerin *et al*. (1996). Uchida and Sasaki (2006) were using MT to unravel the geothermal system in Ogiri area Japanand de Lugao et al. (2002) also applied MT in Cadas Novas geothermal area in Brazil. The authors indicated that the controlled-source audiomagnetotelluric (CSAMT) method provides a moreefficient efforts of identifying the shallow resistivity pattern above a measured geothermal system. Then, Pellerin et al. (1996) used 3D numerical models to evaluate four EM techniques, magnetotellurics (MT), controlled-source audio magnetotellurics (CSAMT), long-offset time-domain EM (LOTEM), and short-offset time-domain EM (TEM), in terms of geothermal exploration. One of their results was combination of CSAMT and TEM measurements can give them the best way to delineate the clay cap in a geothermal system. One example application MT application in Indonesian good examples was brought up by Sumintadireja *et al.* (2010). Using the MT survey, the mentioned author has successfully indicated low susceptibility areas, which related with surface manifestation of hot springs in Jaboi area (Aceh Area). Another example was applied in Mataloko area (Flores, Indonesia), by Takahashi *et al.* (2000) and Sitorus and Aswin (2002) in their unpublished report.

All of the papers have shown the success of MT application to view the subsurface structure of geothermal area in strato volcano. The combined stratigraphy of effusive and pyroclastic materials with contrast of resistivity can be detected nicely with MT survey. Mount Ciremai is also considered as strato volcano (Situmorang, 1995). Therefore, we can expect similar success in identifying prospect area with these measurements.

Most volcanic areas with fresh andesitic breccia layers, lavas, and altered rocks due to the high temperature, will show complex distribution over large area. There fore we need remote sensing analysis as tools to make list of regional prospect zones. Once the prospect zones are identified, then we will need more detail tools to study the geothermal system of the area. MT is one of the useful techniques to do the task. In volcanic areas, Breccia and lava flow often-identified in form of ridges, which fill previous valleys. The distributions also often show differences in between flanks. In case of Mount Ciremai, the west flank contains more pyroclastic material, while the east flank contains dominant lava. The conductivity or resistivity contrast of the two rocks can be detected with MT measurement. According to the above discussion, we can see that the remote sensing and MT method are two useful method to be combined, each method for regional and detail scale.

**2 Previous Studies**

Mount Ciremai is one of active strato-volcanoes well as one the highest volcanoes in Indonesia. Its peak is reaching to about 3078 masl. The volcanic deposits found in the area were the product of four previous eruptions as also described in Situmorang (1995). The long history of eruption has produced large volume of volcanic deposits, which can be classified as: lavas, pyroclastic, and lahars, as shown in the following Figure 2 (Situmorang, 1995). The vast number of volcanic deposit layers has become the active constraint for hydrogeological condition the area.

The rough morphology on this area can be classified as follows based on the slope gradient: 5-10$^o$, 10-25$^o$, 0p-25-40$^o$, consists of highlands with the elevation ranges from 250 to 3000 masl and low lands located at 100-250 masl. Difference in slope gradient is mainly controlled by the change of lithology. Hard and compact rocks such as lavas and pyroclastic are likely tobuild steep slope gradient, while softer rocks, such as lahars, are tend to build more easyslope (Irawan *et al.* 2009). The morphology itself gives substantial control to groundwater flow system. Irawan *et al.* (2003) proposed a regional overview that the hydrogeology of Mount Ciremai is composed by three zones: Zone 1 at 150-250 masl, Zone 2 at 250-750 masl (largest number of spring emergences), and Zone 3 at 750-1250 masl. Large springs, which mostly located at 250 – 750 masl of elevation, are likely controlled by slope break, whereas the slope break itself is formed by the change of rock layers, from steeper slope of lava at higher elevation to easier slope of lahars at lower slope. The direction of main geological structure at Mount Ciremai is almost NW-SE which dissected the Young Ciremai Volcanic products. However, there are also west-east trending fault as a lineament of old Ciremai caldera resulting Sangkanhurip and Pejambon geothermal prospect in the east part of the volcano.

The initial geothermal information of Mount Ciremai and adjacent area is collected by volcanology survey of Indonesia and PT. Pertamina based on field survey result, which was carried out in 1986-1990. The collected data consists of geological map, geomagnetic, geoelectric, shallow drilling temperature, and geochemical data. There are several surface geothermal manifestations surrounding Mount Ciremai, in form of fumarole and solfatara. They were located at the crater with average temperature about $61^{o}C$ and hot springs at volcanic toe with temperature ranging from $33$-$50^{o}C$. Based on their unpublished findings, PT. Pertamina has already pronounced two prospect areas of Mount Ciremai.

Another researcher, Irawan *et al.* (2009), has analyzed the water chemistry on 119 spring locations to identify the possible volcanic aquifer system of this volcano. The author proposed three groundwater clusters based on cluster analysis on 14 parameters. Cluster 1 (112 springs) has normal temperatures, low TDS, EC, and high bicarbonate concentrations. Cluster 2 (five springs) has moderately high temperature, TDS, EC, and high concentration of chloride. Cluster 3 (two springs) exhibits high temperature, anomalous high TDS, EC, and chloride concentration. Each cluster is a resemblance of hydrogeological system. The 1$^{st}$ system is developed in shallow unconfined aquifer, with domination of high bicarbonate (4.2 me/L) meteoric water. The 2$^{nd}$ system is predominated with mixing processes, between unconfined groundwater and hot groundwater from deeper aquifers. The 3$^{rd}$ system is primarily dominated by groundwater flow from deeper formation. The hot – deep seated groundwater flow also carries mud particles. It has anomalous high TDS (>1000 mg/L), EC (515 μS/cm), and chloride (99 me/L) from interaction with clay formations of Kaliwangu Formation.

The Ministry of Earth and Mineral Resources ESDM in 2010 has also made a study on the hydrochemistry. The results were coherent with the previous mentioned research, in which the water from geothermal reservoir mixed with meteoric water and originated from single heat source. These phenomena are occurred at all hot springs this area. One of the main findings was the influence of volcanic magmatic processes, characterized by higher Cl concentration compared to total of B and Li. Hot springs with higher Cl content than B and Li, were influenced by volcanic magmatic processes. Moreover, this project was also measured the silica geothermometer and came with reservoir temperature of 140-180 $^{o}$C.

## 3 Methods

### 3.1 Remote Sensing Analysis

Landsat Thematic Mapper (TM) is a multispectral scanning radiometer that was carried on board Landsats 4 and 5 satellites. Moreover, The Landsat Enhanced Thematic Mapper (ETM) was introduced in the Landsat 7 satellite. ETM data cover the visible, near-infrared, shortwave, and thermal infrared spectral bands of the electromagnetic spectrum. This project is collaboration between the U.S. Geological Survey (USGS) and the National Aeronautics and Space Administration (NASA).

We used four scenes of Landsat ETM+ fromUSGS Global Visualization Viewer from USGS official website (http://glovis.usgs.gov/) with same path and row. The range of acquisition date is in 1999 to 2003. After 2003, Landsat ETM+ could not acquire the data completely due to the failure of Scene Line Corrector (SLC) instrument permanently. Therefore, we limited the data used to only before 2003. Table 1 below shows the data in details.

One of the common image analyses is Land surface temperature (LST). It is an important analysis for a wide variety of applications such as: hydrological, agricultural, biogeochemical and climate change (Van *et al.* 2009). It is strongly influenced by the ability of the surface emission. Therefore, calculating LST will help us to locate the anomalous surface temperature that presumably related to geothermal manifestation. The identification of these anomalies has been applied in a given geothermal area as similar markers of geothermal potential. We also conducted atmospheric correction in the LST analysis using Fast Line-of-Sight Atmospheric Analysis of Spectral Hypercubes (FLAASH) technique for visible to short infrared bands, and single band correction technique for thermal infrared band. Both corrections were applied in order to reduce the noise of atmospheric disturbances from small particles or water vapor in the atmosphere. The noise will attenuate and distort the emitted electromagnetic wave leaving a given location target on the ground.

To calculate the LST, the Landsat ETM+ band 6 (10.4-12.5 μm) in digital number (*DN*) was transformed to radiance as follows:

$$L_\alpha = \frac{L_{max} - L_{min}}{QCAL_{max} - QCAL_{min}}(DN - QCAL_{min}) + L_{min} \qquad (1)$$

Where *Lα* is radiance measured by satellite in W/m2*sr*μm, *QCALmax* and *QCALmin* are the *DN* maximum (=255) and minimum (=0), *Lmax* and *Lmin* are scaled radiant to the *QCALmax* dan *QCALmin*, respectively. After the data in radiance, space-reaching radiance to a surface-leaving radiance can be calculated as follows:

$$L_T = \frac{L_\lambda - L_\mu - \tau(1-\varepsilon)L_d}{\tau\varepsilon} \qquad (2)$$

Where *LT* is the radiance of a blackbody target of kinetic temperature *T*, *Lμ* is the atmospheric path radiance, *Ld* is the sky radiance, *τ* is atmospheric transmission, and *ε* is emissivity of object at surface. Physical parameters of the atmosphere (*Lμ, Ld, τ*) are based on radiance transfer modeling of MODTRAN. We obtained *ε* from the USGS emissivity library (Modis Emissivity Library, available at http://g.icess.ucsb.edu/modis/EMIS/html/soil.html) for sand and soil. Then, the brightness temperature *Tb* could be calculated by following equation:

$$T_b = \frac{C_1}{\ln\left(\frac{C_2}{L_T} + 1\right)} \qquad (3)$$

Where *Tb* is brightness temperature in Kelvin (K), *C1* and *C2* are calibration constant 1 and 2, respectively. Chandler *et al.* (2009) had proposed this approach. The *Tb* is the temperature necessary for a blackbody to emit energy at the same rate as that observed from a gray body (Artis and Carnahan, 1982). Therefore, a correction for spectral emissivity *ε* in eq. 2 is necessary to calculate the *LST* as follows:

$$LST = \frac{T_b}{1 + \left(\lambda + \frac{T_b \times \sigma}{h \times c}\right)\ln\varepsilon} \qquad (4)$$

Where λ is the wavelength of emitted radiance, *σ* is Stefan Boltzman constant, *c* is light velocity, and *h* is Planck's constant.

In order to estimate the influence of temperature from background, a Normalized DifferentVegetation Index (NDVI) for near infrared (NIR) and red bands is used as follows:

$$NDVI = \frac{r_4 - r_3}{r_4 + r_3} \quad (5)$$

Where *r3* and *r4* are reflectance of band 3 and 4, respectively. The purpose of this technique is to exclude the temperature, which has no correlation with surface manifestation in geothermal system.

A Temporal Temperature Mapping (TTM) of LST was used to estimate spatially the distribution of surface temperature anomaly in relation with the surface manifestation of a geothermal system. A field temperature measurement at seven locations and the surface temperature at the crater were used as references for the TTM calculation. The assumption in this method was that the temporal variation of temperature at surface from the geothermal system is more stable over time than temperature from the background. Therefore, the temperature variation at location of surface manifestation can be used as reference *Tr* to estimate the target *Tt*. The similarity of *Tt* and *Tr* are presented by cosine angle *β* as follows:

$$\beta = \cos^{-1} \frac{\sum_{i=1}^{4} T_t T_r}{\sqrt{\sum_{i=1}^{4} T_t^2 \sum_{i=1}^{4} T_r^2}} \quad (6)$$

Where *Tt* and *Tr* are temperature of target and reference, respectively. The reference of temperature is listed in table 2.

**3.2 Magneto-telluric (MT)**

The major problem confronting geothermal exploration is how to efficiently, economically and effectively predict the optimal site for drill holes in order to provide the best chance of intersecting productive thermal fluid channels and reservoirs deep beneath the subsurface. The occurring problem needs more reliable subsurface detection method, and for the time being, geophysical method is still the most widely used.

Geophysical methods to measure sub surface's electrical resistivity properties are basically divided into electrical potential difference and electromagnetic field measures. Both are sharing the same approach, which is comparing natural values to artificial one. Electrical potential measurement has been used mainly for shallow depths, for example for groundwater aquifers or very shallow geothermal reservoirs, whereas the magneto-telluric (MT) method has been used for greater depths. It measures the earth's impedance to naturally occurring electromagnetic wave. Recent development has chosen MT, as the standard in most areas for reconnaissance and detailing purposes in geothermal exploration, because of its simplicity factor. It has been generally used to detect the bottom clay (cap rock) as borehole target and also to predict faults. This method requires sophisticated instrumentation that offers a vast spectrum of possible applications. The main advantage is that it can be used to define deeper structures.

The INVariant-mode, TE-mode and TM-mode were used as the default technique, from the MT equipment software package. The inversion technique, Occam Model and manual model came as the standard technique in the MT software package. Huang and Albrecht (2011) also used the Occam 2-D mathematical inversions were performed along each BMT profile in order to derive a model that represents or fits the data. This modeling allows for a geological interpretation of the geophysical data. The mentioned steps were proposed by Pedersen and Engels (2005), and were used for all 2-D inversions. The procedures invert the apparent resistivity and phase of the determinant of the impedance tensor. The expected results were inversion models with fourth iteration and RMS error of 1. The MT modeling used *Winglinkv2.18* software that came with the instrument.

There were 22 MT measurements, encircling the study area, which were divided into two lines. Line A was consist of 10 points data and Line B consist of 12 points. The interval between each point was 600 - 2000 km. Figure 3 shows detail distribution of the measurements. The sounding 1D modeling was processed based on INVariant-mode, its average of TE-mode and TM-mode. The model was built based on two methods. First method was the Occam Model Inversion as presented in Figure 5 (45 layer maximum, violet color) and the second method was manual model inversion (8 layer maximum, green color).

**4 Results**

**4.1 Remote Sensing Analysis**

**4.1.1 Land Surface Temperature**

Based on TTM method, there are six locations with temperature anomaly from eight classified area as shown in Figure 4. Each anomaly is associated with letters A-F. Different color class indicated similarity of temperature from different reference point. In general the anomalous location A to D consists of 2-3 classes. However, location E and F are consisted of one class only. Their locations are situated at lineament pattern of southern mountain range.

The surface temperature can be originated not only from geothermal system, but also from background, such as: human activity, topographic slope orientation (Eneva et. al., 2007), albedo, and thermal inertia (Coolbaugh *et al.* 2003). Among those factors, the human activity is supposed to contribute the highest influence due to the existence of urban area around the Mount Ciremai. For such the reason, we used the NDVI with certain threshold to delineate the possibility of temperature anomaly from human activity. The criterion of delineation is based on the density of vegetation in which rare vegetation may have correspondence with urban area. In general case, the NDVI has negative correlation with the LST. Figure 5shows the temperature anomaly overlaid on the vegetation map. The A, B, and C zones are located on the rare vegetation. It can be interpreted that the human infrastructure such as building, road, and pavement may contribute to the classification result. However, based on field survey, the surface manifestation located in A zone. Therefore, the classified zones are the sum up of all factors.

The D, E, and F zones are situated at dense vegetation. Therefore, the TTM anomalies from human activity are weak. The possibility of temperature from geothermal system should be high at those zones. Some part of D and F zones located at non-vegetated area. The temperature anomaly from those parts may originate from albedo of bare land or water surface. In this meaning, the D and F zones are also not free from disturbance of external factor. The E zone is located at rare vegetation around the mountain peak. High elevation is the cause of the condition, rather than no human activity. The possibility of bare land also low due to existence of vegetation rarely. Thus, the anomaly should be in relation with temperature from crater or up flow zone.

The ambiguity of temperature anomaly could not be detached because many factors contribute to the surface temperature detection. However, it can be reduced and interpreted based on the field knowledge related to the mentioned condition.

**4.1.2 Anomaly of surface temperature**

A Temporal Temperature Mapping (TTM) of LST was used to estimate the distribution of surface temperature anomaly in relation with the surface manifestation in a geothermal system. A field temperature measurement at seven locations and the surface temperature at the crater were used as references of TTM calculation. Figure 6 shows the location of surface manifestation in green dots. Table 2 is the list of the field temperature measurement at different location as shown by Figure 2 and labeled by letter C1 to C7.

The temporal variation of temperature at eight surface manifestations is depicted in Figure 3. Generally, temperature at each location is decreasing, except at point C7 which the temperature is decreasing from Nov. 5, 1999 to May 24, 2002 and then increasing until Jan 19, 2003.

## 4.2 Geophysical Analysis

### 4.2.1 Magneto-telluric (MT)

A coherency analysis was applied to investigate the potential of remote reference processing. Remote Reference is a technique utilized to account for cultural electrical noise by acquiring simultaneous data at more than one MT station. This step can improve data quality, and also allow data acquisition in areas with difficulties to detect natural MT signal because of man-made EM interference. In general, remote referencing yields best results to reject error elements from final stacking, in particular for the period range of 1 to 10 s. This was a crucial step in data improvement and resulted in more consistent apparent resistivity curve with respect to the phase curve. Then, estimation of the impedance tensor from the spectra was done by statistical means. Both allow for remote referencing to reduce the influence of uncorrelated noise. Signal processing has been carried out for all frequencies in general with detail process on LF2 band since information on the target depth is contained in this band.

The T data has been modeled using *Winglink v2.18*. The sounding 1D modeling was processed based on INVariant-mode, its average of TE-mode and TM-mode. The model was built based on two inversion techniques. First method was the Occam Model Inversion as presented in Figure 5 (45 layer maximum, violet color) and the second method was manual model inversion (8 layer maximum, green color) (Figure 8).

The 2D cross section on Line A was constructed using the software, based on the 1D model inversion result. The modeling was based on Bostick Smooth Model data and also Occam Smooth Model Inversion (Figure 9 and Figure 10). From both figures and also considering other subsurface data, we interpret the red color as volcanic zone or in this case is volcanic core of Mount Ciremai, and the green to soft blue color as bedrock zone, which is lava layers. However the prospect zones are in dark blue color, which is more porous or more fractured layers of laharic breccias or pyroclastics. Our complete interpretations are in the later figures.

The 2D model was based on two-inversion method, Smooth Inversion Model and Sharp Boundary Inversion Model (Figure 8 and 9). Both models used Root Mean Square (RMS) error lower than 5%. Line A and B section have been processed using the same method. Both models show similar results, which are volcanic zone (red color) more on the left side and bedrock zone (green to soft blue) below the volcanic zone and slightly undulated on the right side of the section. The prospect zone (dark blue color) lies on the middle and shallow part of the section.

## 5 Conclusion

The remote sensing and MT results demonstrated a good data match for revealing subsurface condition in geothermal area. Remote sensing analysis is put in the position of capturing the potential geothermal area along with the classification. The MT survey will refine and pin point the detail geological system beneath the surface that eventually will delineate the geothermal system. An atmospheric correction in the remote sensing analysis is effective for sub-pixel analysis because it removed the atmospheric disturbance in the data. The combination of temperature anomaly from TTM and NDVI map can be used to estimate the temperature contribution from human activity. Based on TTM classification, there are six anomalous zones: three zones located at northern flank, two zones located at southern flank, and one zone located at western flank of Mount Ciremai. Several factors could contribute to the temperature anomaly, such as human activity at northern zone due to existence of urban area and emissivity from bare soil at southern zone. Then MT interpretation has led to the conclusion that the temperature anomaly at east area is derived from faults. There are four buried normal faults in the area, which positively is part of the conduits in the geothermal system east of Mount Ciremai. Each of the MT section shows the locations of volcanic zone, bedrock zone, and the prospect zone.

**Acknowledgements**

We would like to convey our sincere gratitude to the management of Applied Geology Research Division, Faculty of Earth Sciences and Technology, ITB for supporting this research. We also thank to The Agency of Energy and Mineral Resources of West Java Province for funding this research. We dedicate also our gratitude to the University Partnership Program (UPP) Chevron Indonesia Company for full financial support to attend the 36thWorkshop on Geothermal Reservoir Engineering in Stanford University, on January 31 - February 2, 2011. We express our appreciation to the two anonymous reviewers for their critical comments and valuable suggestions improving the clarity of this paper, and our laboratory assistants with their help on artwork editing.

Table 1 List of Landsat ETM+ data used in this study.

| No | Path/Row | Acquisition | |
|---|---|---|---|
| | | Date | Time |
| 1 | 121/65 | 5 Sept. 1999 | 09:47:14 AM |
| 2 | 121/65 | 24 Mei 2002 | 09:42:59 AM |
| 3 | 121/65 | 9 Jun. 2002 | 09:42:51 AM |
| 4 | 121/65 | 19 Jan. 2003 | 09:42:47 AM |

Table 2 Field measurement of Land Surface Temperature (LST) at location of surface manifestation

| Date | No | C1 ($^o$C) | C2 ($^o$C) | C3 ($^o$C) | C4 ($^o$C) | C5 ($^o$C) | C6 ($^o$C) | C7 ($^o$C) | Crater ($^o$C) |
|---|---|---|---|---|---|---|---|---|---|
| Nov. 5, 1999 | 1 | 41.41 | 45.76 | 44.9 | 38.73 | 37.83 | 34.25 | 41.41 | 47.47 |
| May 24, 2002 | 2 | 37.28 | 37.28 | 37.28 | 34.52 | 35.45 | 34.52 | 24.88 | 38.19 |
| Jun. 9. 2002 | 3 | 36.32 | 36.32 | 36.32 | 33.65 | 32.74 | 31.83 | 33.65 | 37.2 |
| Jan. 19, 2003 | 4 | 27.79 | 33.22 | 33.22 | 28.89 | 29.99 | 24.41 | 42.47 | 19.77 |
| Field Measurement | 5 | 34 | 29.2 | - | 28 | 29 | 28.9 | 23.4 | - |

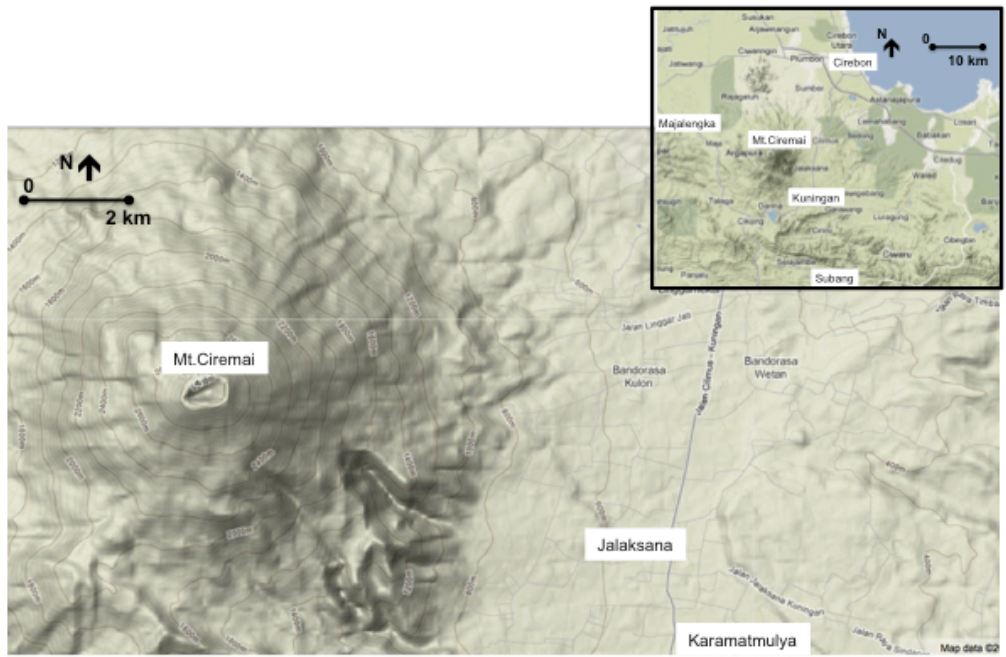

Figure 1 The location of study area, Mount Ciremai, Kuningan Regency, West Java.

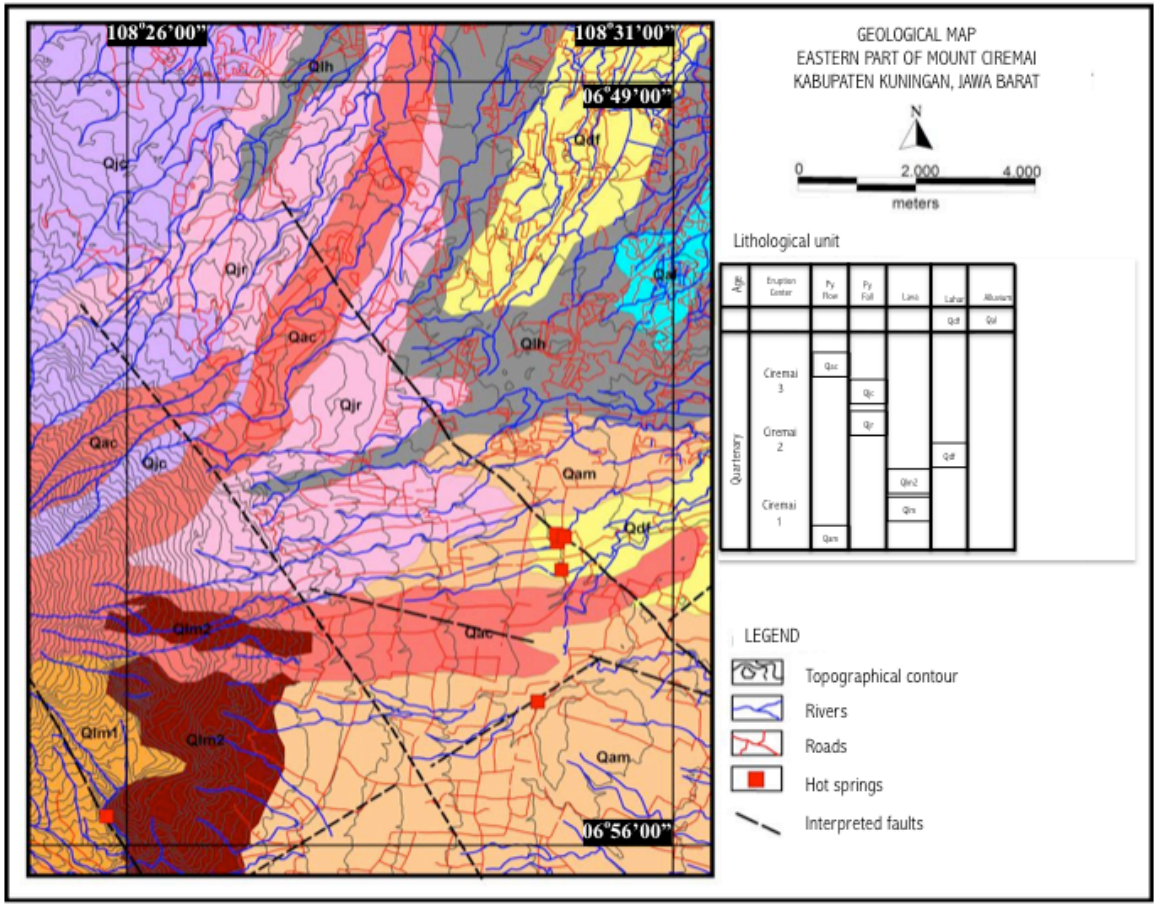

Figure 2 The geological map and N-S and E-W sections of Mount Ciremai (Situmorang, 1995)

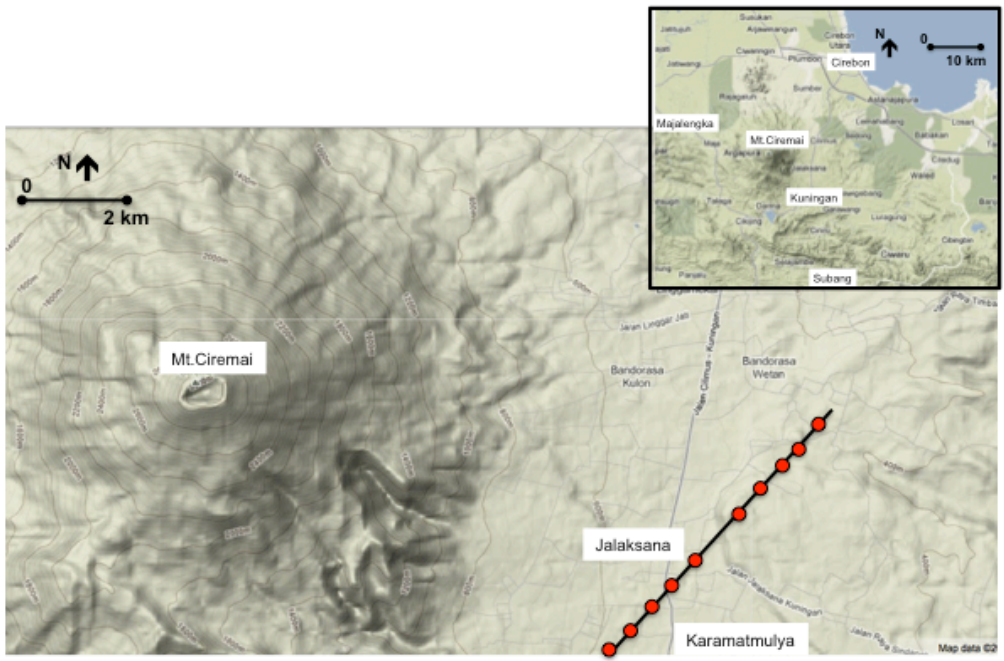

Figure 3 Distribution of MT data and lines

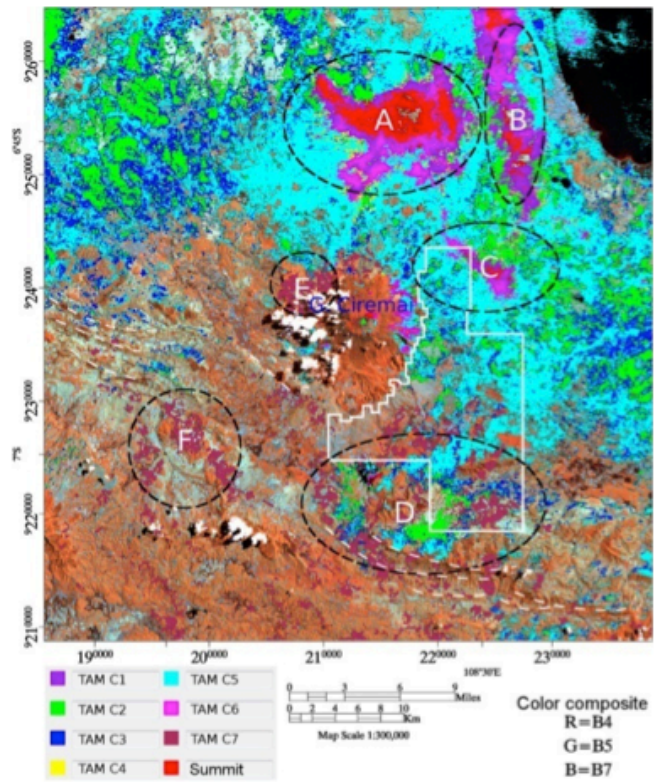

Figure 4 Surface temperature anomaly detected by TTM classification method.

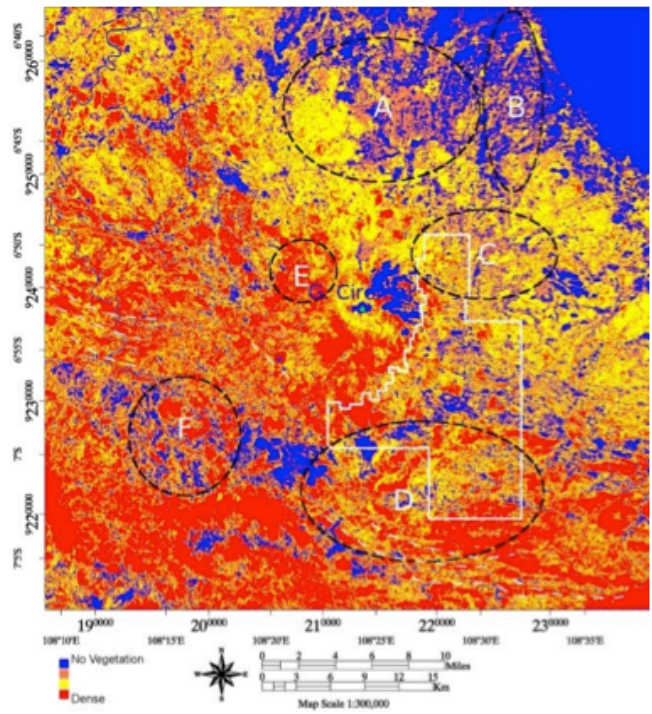

Figure 5 Location of temperature anomaly overlaid on NDVI map.

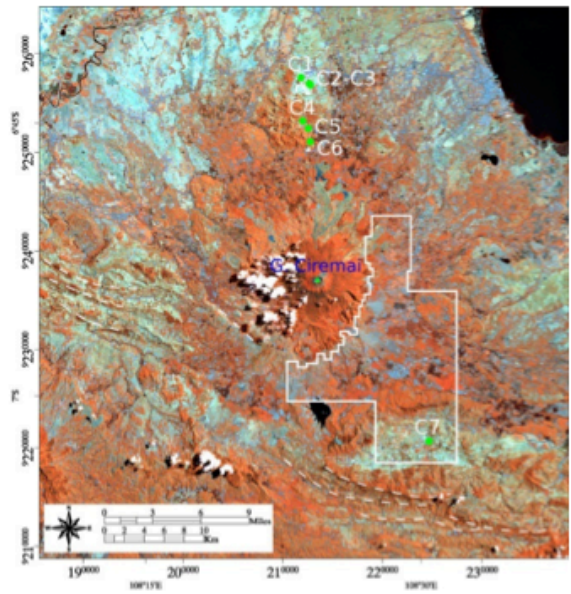

Figure 6 Location of surface manifestation of a geothermal system in green dots with letters C1 - C6 overlaid on Landsat ETM+ image in color composite R, G, B = B and 3, 2, 1, whereas the rectangle is a boundary of exploration.

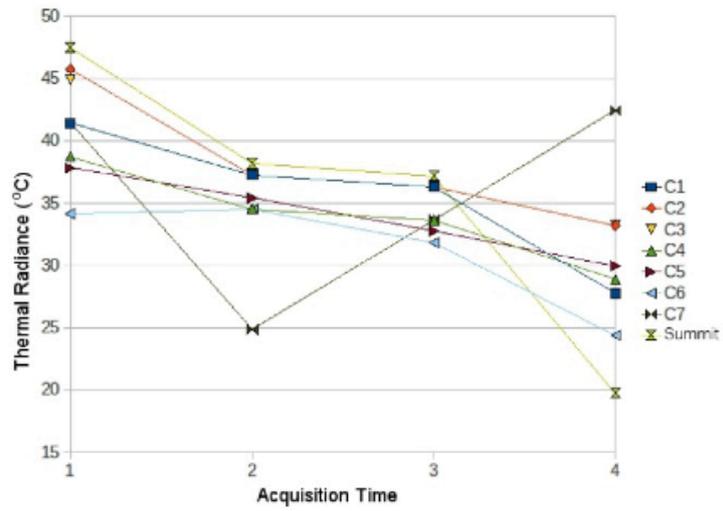

Figure 7 Variation of surface temperature at eight surface manifestations over time. The acquisition time is following Table 2.

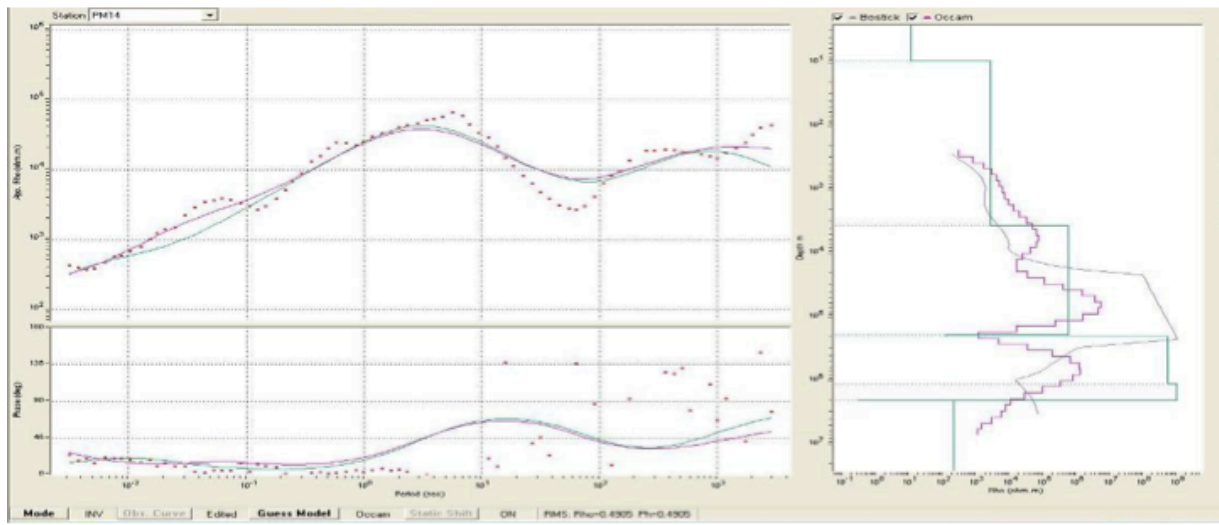

Figure 8 Sampel 1D modeling data based on INVariant-mode using the standard equipment software

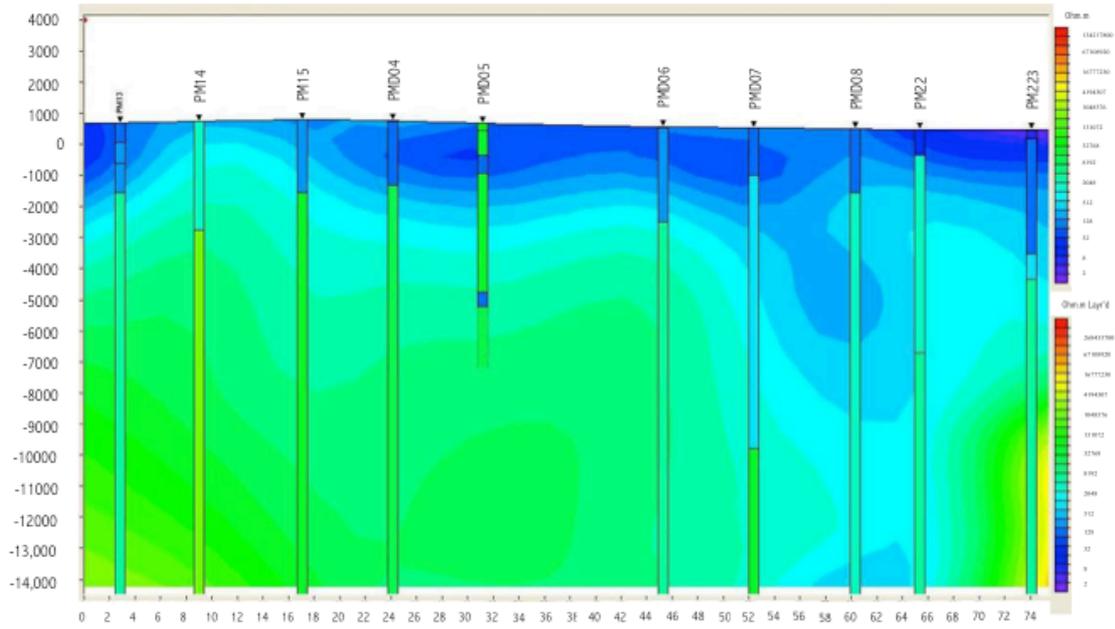

Figure 9 2D cross section from Bostick Smooth Model Inversion method in Line A

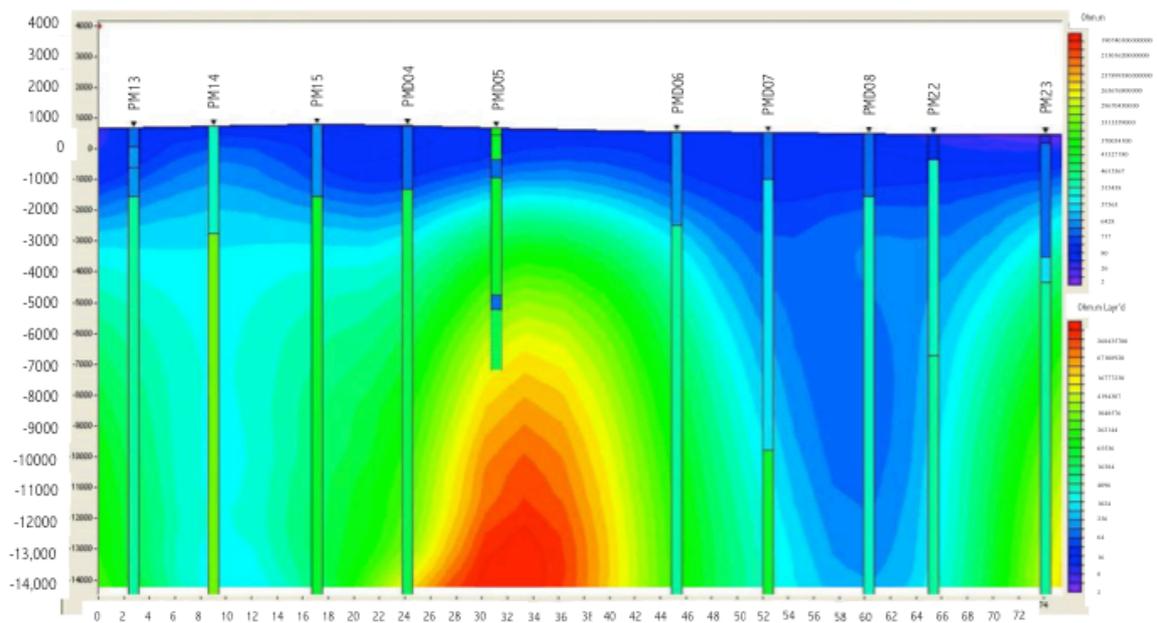

Figure 10 2D cross section from Occam Smooth Model Inversion method in Line A

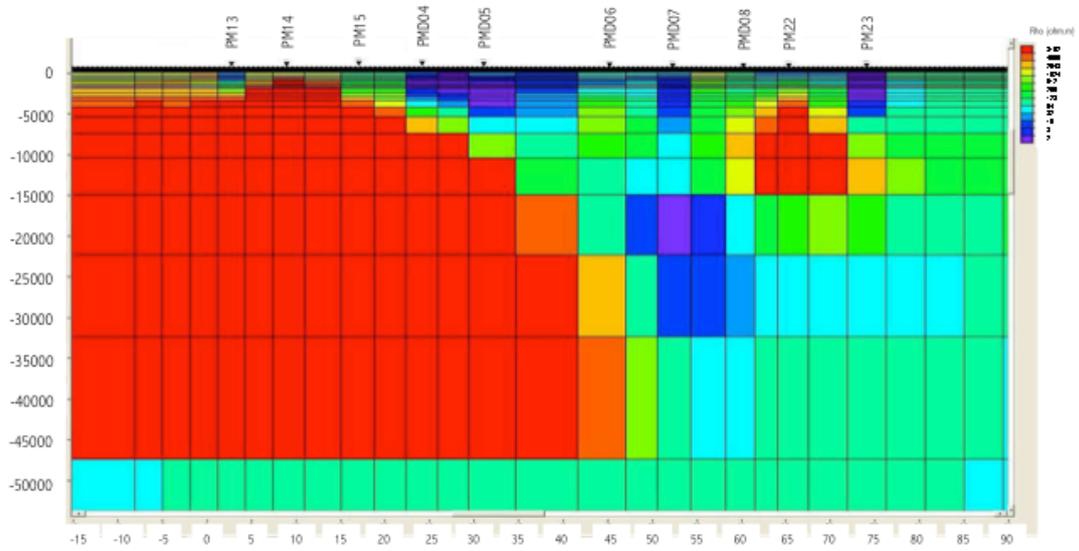

Figure 11 2D-Mesh model from Smooth Inversion Model on LineA

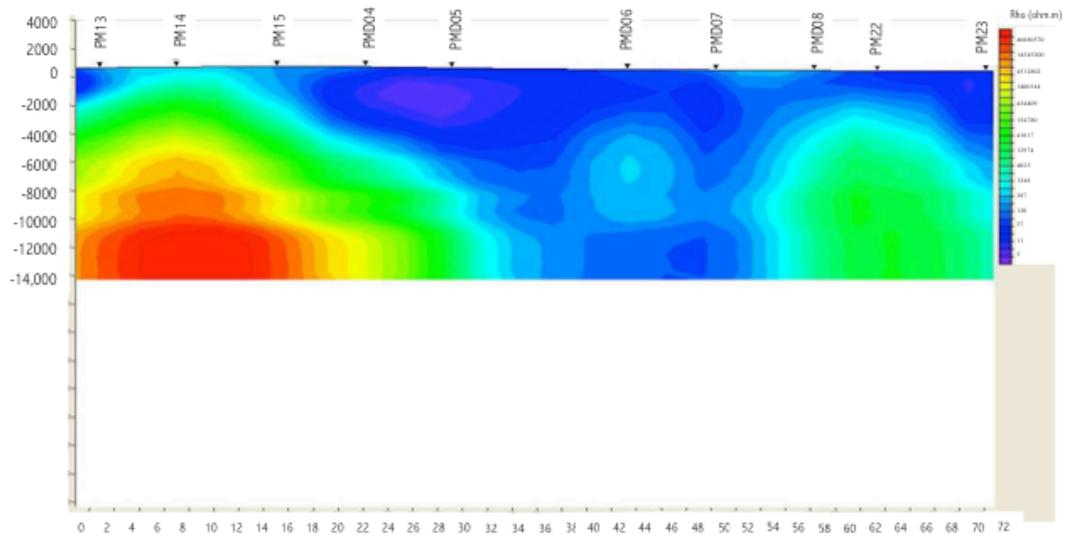

Figure 12 2D-gridded model from Sharp Boundary Inversion model on Line A

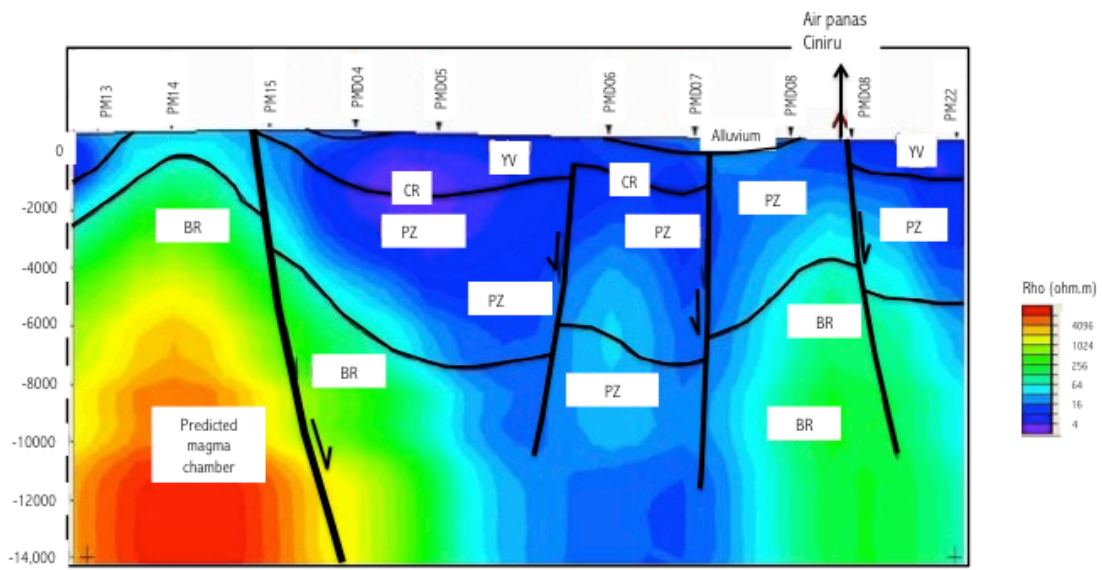

Figure 13 Geological interpretation of MT 2-D with sharp boundary model